# Flexible and Generic Framework for Complex Nuclear Medicine Scanners using FreeCAD/GDML Workbench

Anh Le[1,*], Amirreza Hashemi[2,*], Mark P. Ottensmeyer[3] and Hamid Sabet[4]

[1]Northeastern University, Department of Mechanical Engineering, le.nguyena@northeastern.edu
[2]Massachusetts General Hospital & Harvard Medical School, Department of Radiology, sahashemi@mgh.harvard.edu
[3]Massachusetts General Hospital & Harvard Medical School, Department of Radiology, mottensmeyer@mgh.harvard.edu
[4]Massachusetts General Hospital & Harvard Medical School, Department of Radiology, hsabet@mgh.harvard.edu

*These authors contributed equally to this work.

Corresponding author: Amirreza Hashemi, sahashemi@mgh.harvard.edu

**Abstract.** The design of nuclear imaging scanners is crucial for optimizing detection and imaging processes. While advancements have been made in simplistic, symmetrical modalities, current research is progressing towards more intricate structures, however, the widespread adoption of computer-aided design (CAD) tools for modeling and simulation is still limited. This paper introduces FreeCAD and the GDML Workbench as essential tools for designing and testing complex geometries in nuclear imaging modalities. FreeCAD is a parametric 3D CAD modeler, and GDML is an XML-based language for describing complex geometries in simulations. Their integration streamlines the design and simulation of nuclear medicine scanners, including PET and SPECT scanners. The paper demonstrates their application in creating calibration phantoms and conducting simulations with Geant4, showcasing their precision and versatility in generating sophisticated components for nuclear imaging. The integration of these tools is expected to streamline design processes, enhance efficiency, and facilitate widespread application in the nuclear imaging field.

**Keywords:** Nuclear Imaging, Computer-Aided Designing, Medical Scanner.

## 1 INTRODUCTION

Nuclear imaging plays a vital role in medical diagnostics, providing detailed images of physiological processes within the body through the detection of radioactive tracers. Techniques such as positron emission tomography (PET) [1–6] and single photon emission computed tomography (SPECT) [7–9] have evolved significantly over the years, becoming indispensable tools in both clinical and research settings. Initially, these imaging modalities were characterized by relatively simple and symmetrical designs, which offered sufficient resolution and sensitivity for many applications. However, the need for more precise and detailed imaging has driven the development of increasingly complex structures. Advancements in detector technology, particularly in scintillators and photodetectors, have allowed for improved spatial resolution and sensitivity, enabling the visualization of finer anatomical details and more accurate quantification of tracer distribution. Modern nuclear imaging systems now incorporate intricate geometries and advanced materials to optimize performance, including innovations in detector configurations. In this paper, we introduce the FreeCAD/GDML workbench (WB) to effectively design and test these intricate geometries, with particular focus on nuclear imaging modalities [10,11].

FreeCAD is a versatile, open-source 3-dimensional (3D) CAD modeler designed for various applications. It features parametric modeling, a modular architecture, and integration with several open-source libraries, enhancing its capabilities for complex design and analysis tasks [10]. GDML (Geometry Description Markup Language) is an XML-based language ideal for describing complex geometries for simulations that follow the Geant4 protocols. It enables precise definitions of geometries, simulation compatibility, and the creation of nested structures, which are essential for designing detector arrays and complex assemblies [10]. The GDML WB is a FreeCAD add-on for exporting FreeCAD geometries to a GDML file, and for importing a GDML file into FreeCAD. The GDML WB is not bundled with FreeCAD but must be installed via the FreeCAD Addon Manager [11]. This process involves selecting the GDML WB from the available addons, which integrates it into FreeCAD. We note that the GDML's WB tools can be combined with other FreeCAD preinstalled or addon WB's, (e.g. Part for 3D modeling, and Body for managing solid bodies) allowing for versatile and complex designs. In addition to exporting the standard GDML solids, the WB allows the export of FreeCAD arrays, extrusions, revolves, and reflections or even changes in properties following the common protocols of the particle physics Monte Carlo simulation platform, Geant4 [12]. Moreover, the output is compatible with software based on the Geant4 engine such as VPG4 [13,14]. Following the Geant4 protocol is a contrasting feature of the FreeCAD/GDML WB, which sets it apart from other CAD platforms. Additionally, FreeCAD Macros (python scripts) can be used to create user-specific tasks in cases where the standard FreeCAD array structures are insufficient. Therefore, it offers users substantial benefits in terms of automation, quick geometry generation, and efficiency. Important features of the GDML WB that are specifically relevant to medical imaging scanners include:

- Polar and Planar Arrays: Facilitate the creation of regularly spaced detector arrays, crucial for PET and SPECT scanner designs.
- Tessellated Meshing: Enables the creation of intricate, mesh-based structures, improving the fidelity of simulated geometries.
- Mirror Copy/Displacement: Simplifies the replication and positioning of symmetrical components, ensuring accurate and efficient design of scanner elements.

Given the capabilities of FreeCAD and the GDML WB, along with the current state of nuclear imaging, this paper integrates and provides the necessary tools to facilitate the design and simulation of PET and SPECT scanners. In the following sections, we will introduce the design principles of PET and SPECT systems, discuss the creation of phantoms for calibration, and demonstrate the application of FreeCAD/GDML WB in these processes. Additionally, we will illustrate both the usefulness of FreeCAD/GDML WB and the flexible import and export of GDML files in simulations with Geant4. Ultimately this paper aims to pave the way for easy-access applications to generate complex components for nuclear imaging modalities.

## 2 POSIRON EMISSION TOMOGRAPHY (PET)

PET is a high-energy imaging technique that provides high-resolution images of the body's functional processes. PET scanners detect pairs of back-to-back ~511 keV gamma rays emitted when a positron from a decaying radionuclide in the tissue being imaged annihilates with a nearby electron. Their design entails arranging scintillation detectors in a radially symmetric configuration, such as a circular or hexagonal array, that encircles the patient. These detectors convert gamma rays into light, which is then transformed into electrical signals by photomultiplier tubes (PMTs) or silicon photomultipliers (SiPMs). The precision arrangement and synchronization of these detectors enable accurate localization of positron emissions, facilitating high-resolution imaging [15–18]. PET scanner radiation detectors each contain a scintillator crystal made from materials like lutetium oxyorthosilicate (LSO) or lutetium-yttrium oxyorthosilicate (LYSO) to convert gamma rays into visible light photons. Positioned behind each crystal is a PMT or SiPM that converts these photons into amplified electrical signals. A data acquisition system (DAQ) collects and processes these signals, converting them into digital data used for image reconstruction. Some PET scanners include a collimator to focus gamma rays for improved spatial resolution. The gantry, which houses the detector ring and encircles the patient bed, rotates around the patient during imaging. Advanced computer systems and software are essential for image reconstruction, data analysis, and visualization. Radiotracers injected into the patient emit positrons, generating gamma rays that are detected by the PET scanner to produce detailed 3D images depicting metabolic and physiological processes in tissues and organs. To better understand the mechanics and design variations of PET scanners, it is important to explore their structural components and configurations, starting with the simplest design.

### 2.1 Single PET structure- single ring

We begin by creating the fundamental elements of PET. Figure 1 shows the tools to create components of a PET in the FreeCAD/GDML WB with three main components. One is the set of creation tools, which allow users to create the 3D shape of a box, a sphere or a trapezoid. The dimensions, material, and placement (position and rotation) of the created shapes can be edited in the property data panel that FreeCAD displays.

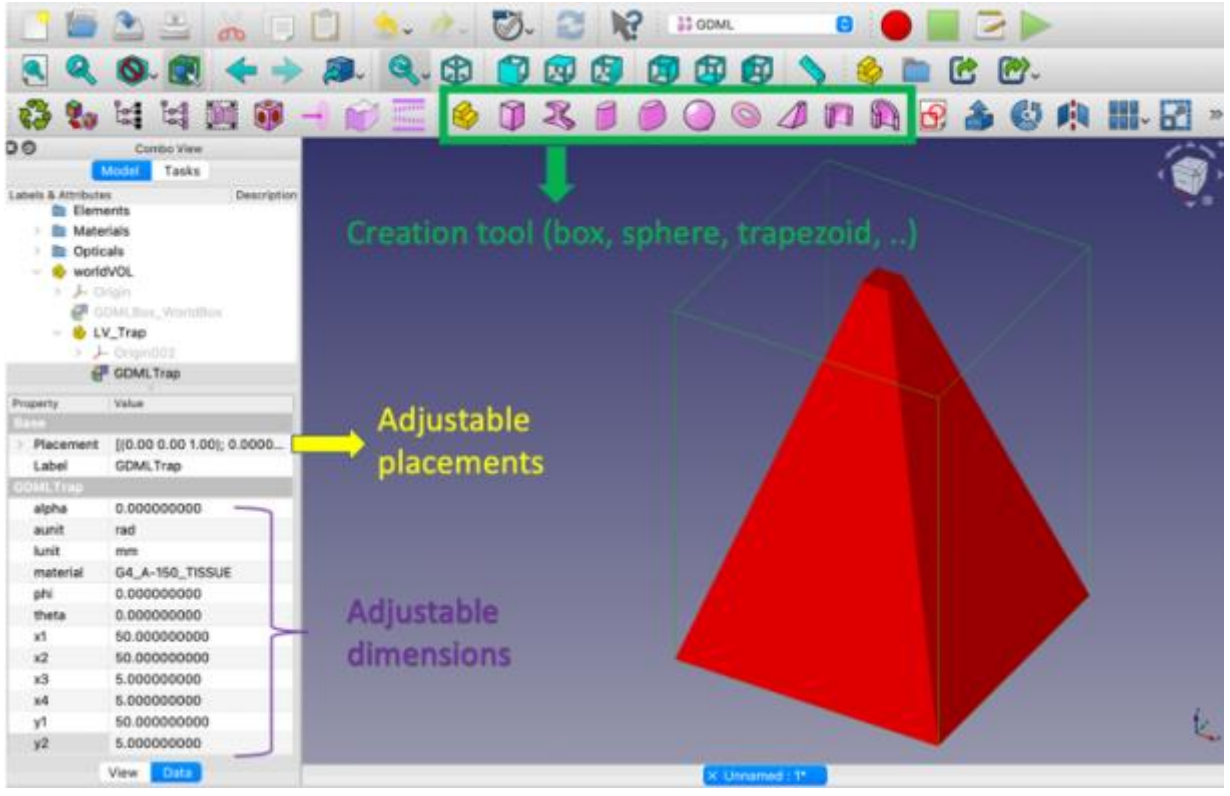

**Figure 1**: Demonstration of the GDML WB layout with the building toolbox for simple geometry creation.

For a single-ring PET, the GDML WB's GDMLBox tool is used to define the foundational elements. For instance, a large box is created as the primary container for the PET components. Simple shapes like spheres and smaller boxes can then be added using the sphere creation tool and additional GDMLBox commands. These shapes are positioned within the primary container by specifying their dimensions, positions, and orientations. The set of tools is indicated in the green box of Figure 1.

Within the GDML WB, one useful tool is the array function, which allows users to replicate basic elements across defined intervals, generating multiple instances arranged in a grid-like pattern. This tool is especially relevant for designing the linearly repeated structures of a PET system. Parameters such as the number of elements and spacing intervals can be adjusted to meet specific design criteria. Figure 2 shows an example of an orthogonal array of 8x8 photomultiplier detectors attached to the scintillator for a single PET detector. The scintillator is a 3D box of dimensions 27 x 27 x 10 mm$^3$, and the photosensors are an 8 x 8 array of 3.2 x 3.2 x 2 mm$^3$ Si photomultiplier (SiPM) crystals.

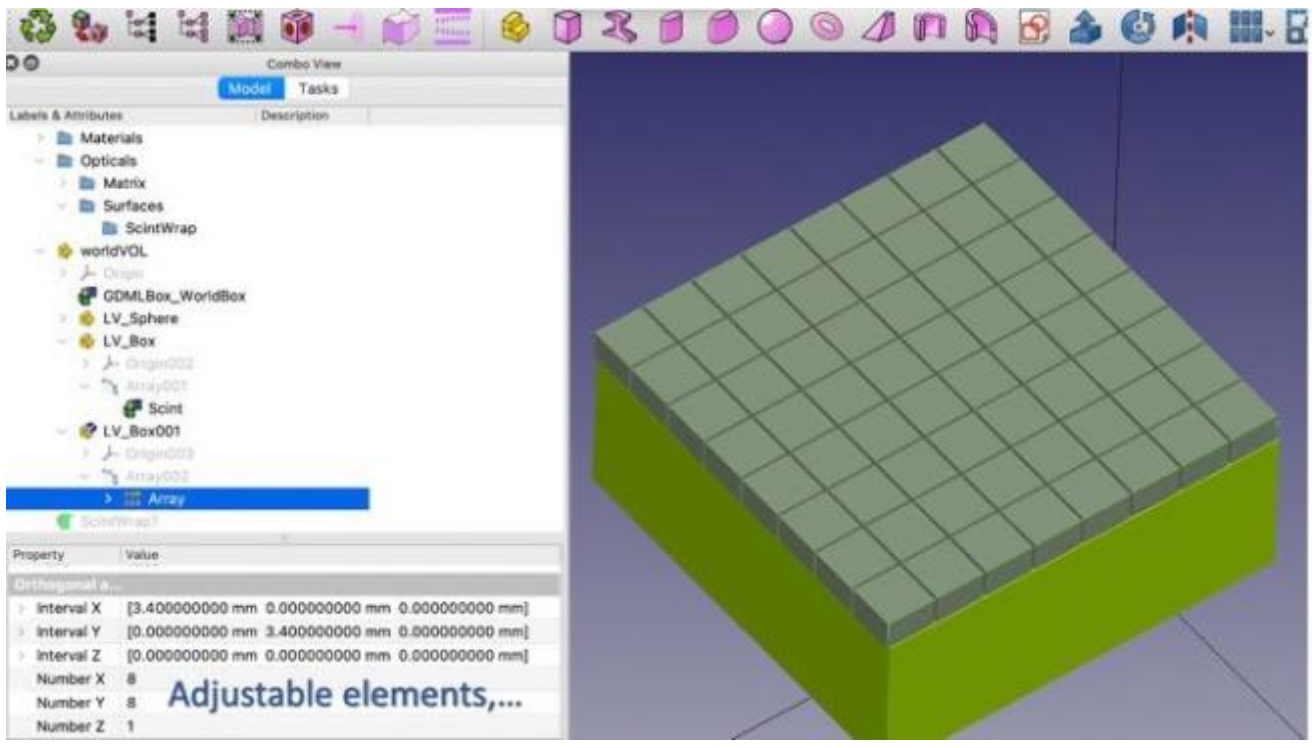

Figure 2: Planar array tool shown in GDML WB.

Furthermore, the Polar Array function simplifies the creation of radially symmetrical rings for PET structures, such as the octagonal structure in Figure 3. By setting the number of elements and positioning them accordingly, a ring is created. Figure 3 presents an example of a polar array of eight elements, each of which is an 8 x 8 orthogonal array of 3.2 x 3.2 x 2 mm$^3$ crystals.

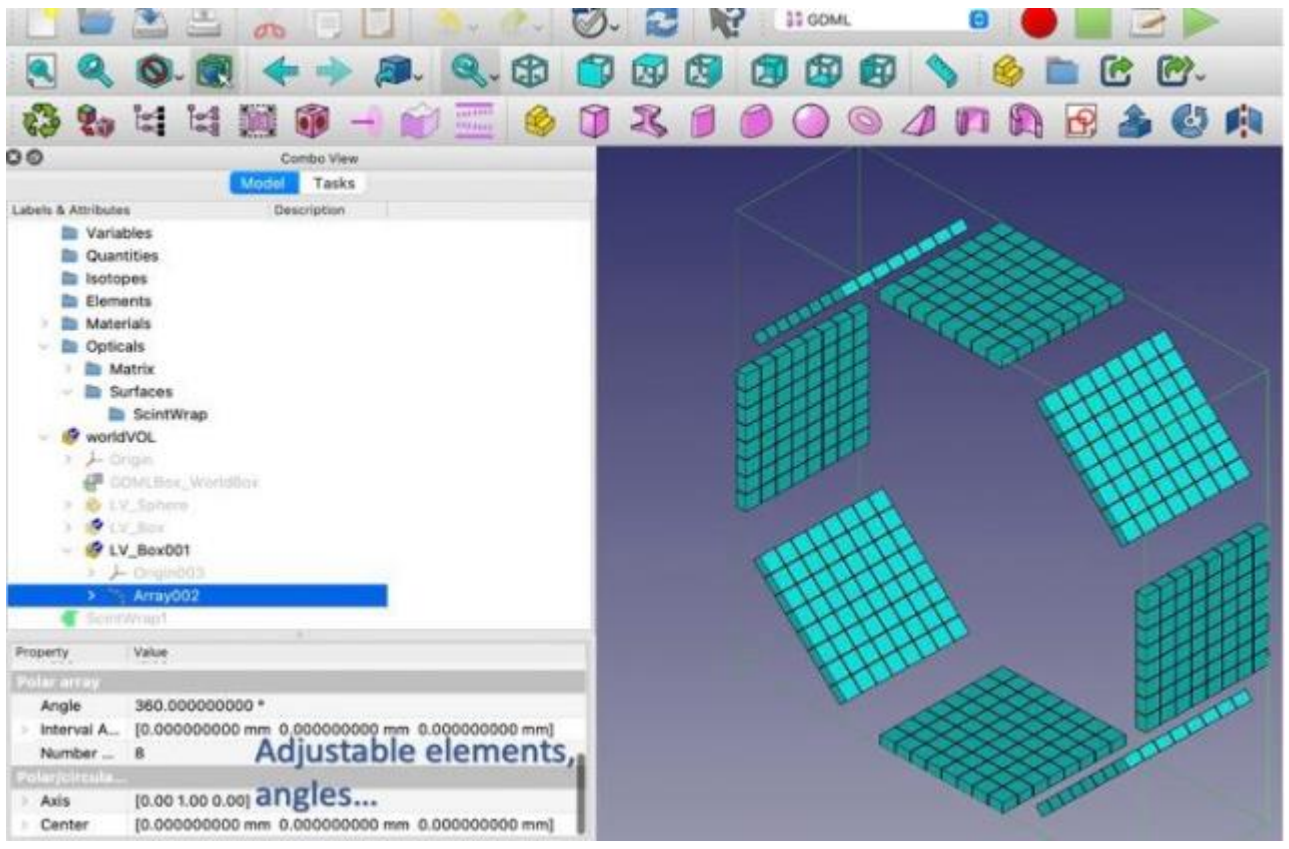

Figure 3: Demonstration of the Polar array tools.

As indicated in Figure 3, arrays can be nested, such as in a polar array of a planar rectangular array of detectors. Beginning with simple shapes and leveraging array functions enables users to quickly implement these techniques, leading to more complex and efficient designs. Figure 4 shows a single ring of a simple PET structure created with a 3D box, sphere creation, and a polar array of a planar array [19].

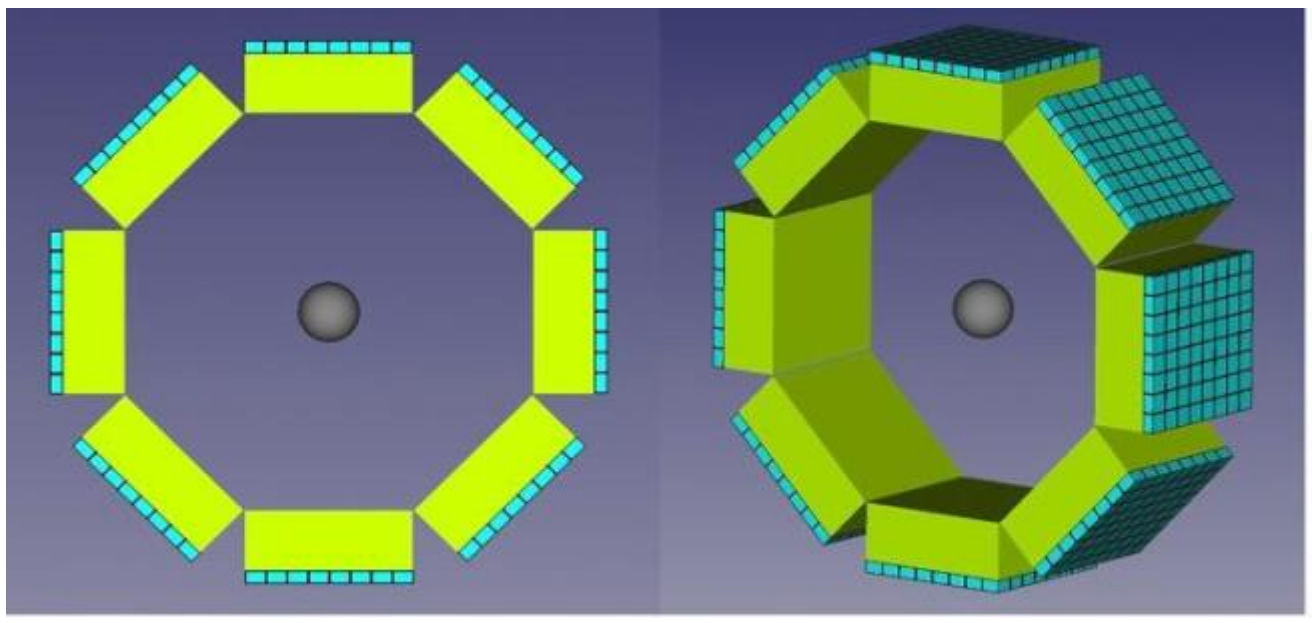

Figure 4: Demonstration of a PET structure–single ring.

## 2.2 Simple PET structure-multiple ring (whole body PET)

Building on the foundational concepts of PET structures, we use FreeCAD/GDML to scale up to larger PET designs, such as a whole-body PET scanner incorporating multiple rings and modifiable ring configurations. In this design, we begin with simple box shapes that are organized into intricate arrays. For instance, creating a 3.2 x 3.2 x 10 $mm^3$ box and arranging it in a 7 by 6 array with specific intervals forms the basis for larger configurations [20–22]. These arrays are further expanded into larger grids and positioned precisely within the 3D space. Using the Union tool, setting the axis, and specifying the number of elements, we can create multiple rings of detectors for full-body scanning. These rings are then replicated to form a series of arrays, which can be adjusted in position and size to meet the design requirements. The rings can be created both by array and copy functionalities. Figure 5 shows an overview of a whole-body PET, which is achieved from a single ring PET that is discussed in the previous section by using planar array based on the number of rings [23].

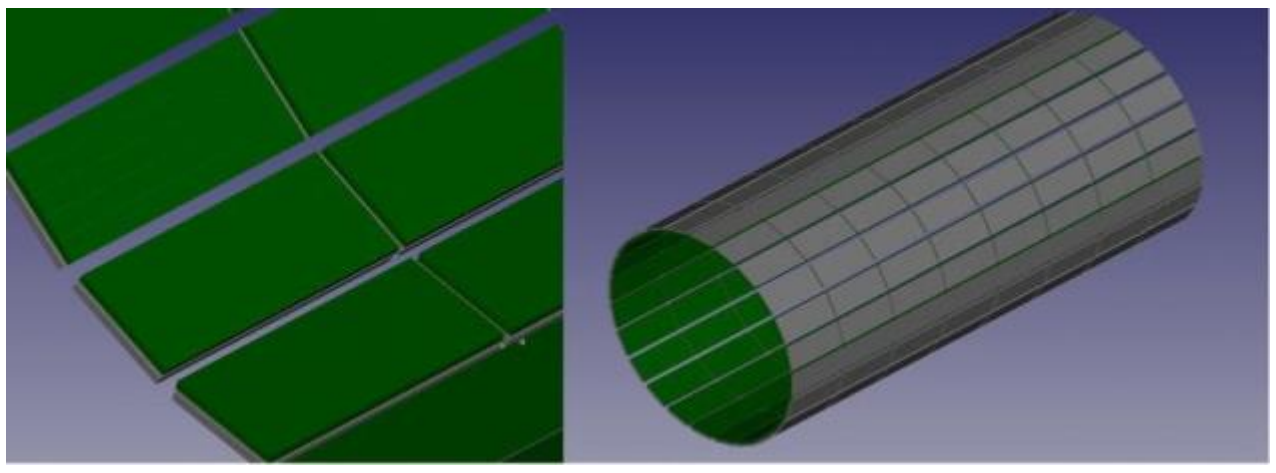

Figure 5: Demonstration example of the PET structure – whole body PET.

Different structural configurations, such as missing rings and missing detectors, can be created by modifying the array parameters. These configurations are particularly relevant to structures with sparse detector positionings. Figure 6 shows missing rings and missing detectors from a whole-body PET example, obtained by the removal of certain elements of the planar and/or polar arrays. The simplicity of the CAD design process enables us to achieve any other variation of a whole-body PET structure quickly.

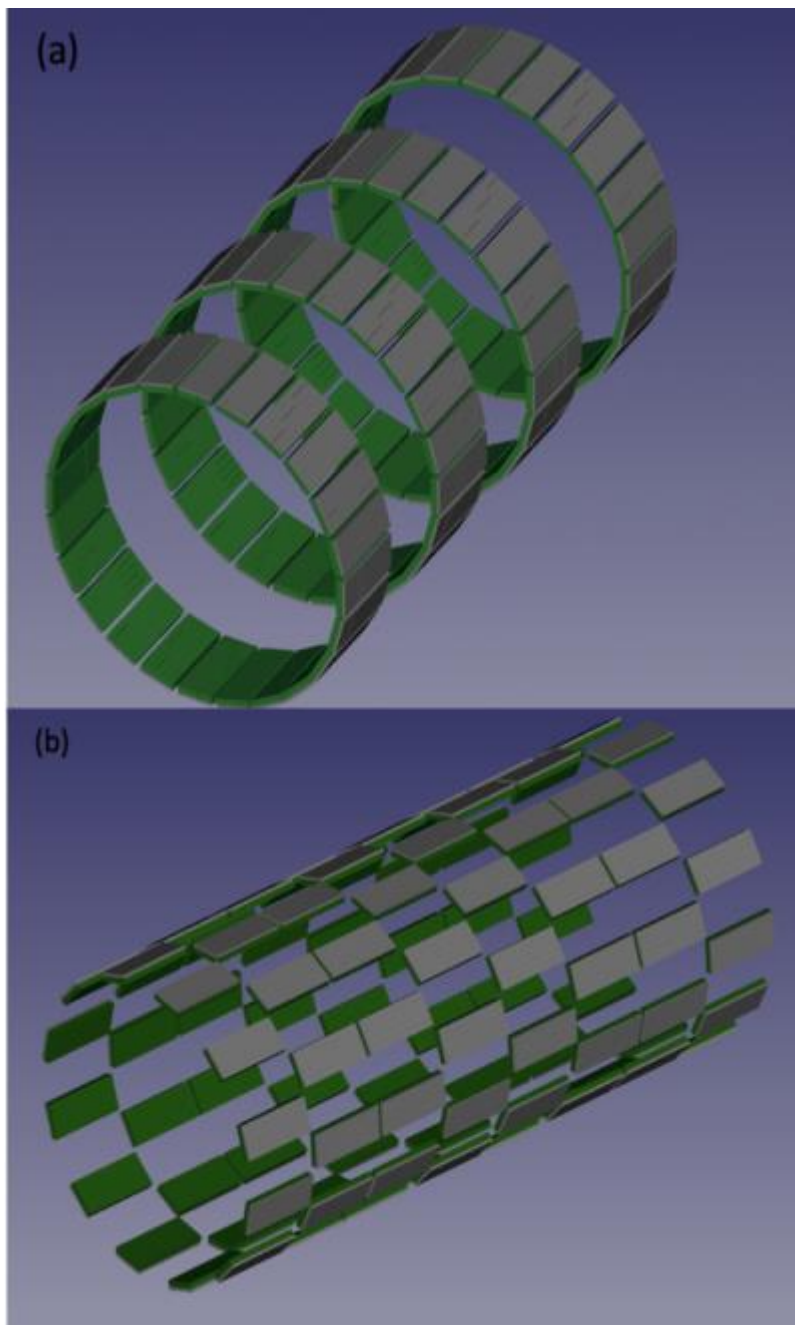

**Figure 6**: Whole body PET with: (a) missing rings; and (b) missing detectors.

## 2.3 Coupled 4 to 1 prism PET & Coupled 9 to 1 prism PET

Prism-PET was developed using a prism-shaped detector to achieve advanced imaging capabilities, including improved resolution and sensitivity [21]. It features segmented prismatoid light-guides with distinct center, edge, and corner designs to ensure uniform crystal identification. By confining light sharing to nearest SiPM neighbors, it enhances the signal-to-background ratio, energy and depth of interaction (DOI) resolutions. The segmentation pattern decouples adjacent crystals, enhancing identification. Right triangular prisms improve inter-crystal light-sharing ratios, further boosting performance. Here, we utilize FreeCAD/GDML to construct a PET structure to showcase the effectiveness and flexibility of these tools in developing intricate and detailed geometries, such as coupling 4 or 9 crystals to a single prism detector. This process uses a combination of basic shapes, arrays, and transformations to build the prism PET components effectively.

We begin with a GDMLBox that is set as the initial structure, providing a base of 120 x 120 x 120 $mm^3$. Next, smaller elements such as boxes and trapezoids are created to represent the crystals and detectors of the PET system. These elements are sized and positioned to fit within the overall structure, with specific configurations like coupling 4 or 9 scintillator crystals to a single prism detector [24–27]. Modules that were examined include one that consisted of a 16 x 16 array of 1.4 x 1.4 x 20 $mm^3$ LYSO crystals coupled 4-to-1 and a second that used 0.96 x 0.96 x 20 $mm^3$ LYSO crystals to achieve 9-to-1 coupling [28–32]. Further details of this structure are available in [21].

Additional components, such as corner detectors and 4 sides & center detectors, are constructed in a similar manner. Unlike the usual box-shaped array, in the coupled 4-to-1 prism case, the detectors involved have different shapes. Corner detectors maintain a box shape, whereas the 4 sides and center detectors feature a box-shaped base with a trapezoidal top. We utilized both box and trapezoid tools to achieve the desired shapes, and then employed the "union" tool to combine all the created parts into a cohesive whole. A similar method was used for the coupled 9-to-1 prism case, with the key difference being that the 4 side detectors are box-shaped while the center detectors have a box-shaped base and a trapezoidal top. These geometries are replicated using ortho array functions to create the complete detector. These components are then positioned and rotated to seamlessly integrate into the PET structure, ensuring alignment and optimal functionality. Throughout this process, elements are adjusted and reoriented to align with the overall design, to achieve the symmetry and efficiency of the PET system. Figures 7 & 8 illustrate the more complex single-ring PET structures, specifically the coupled 4-to-1 prism PET and the coupled 9-to-1 prism PET [33,34].

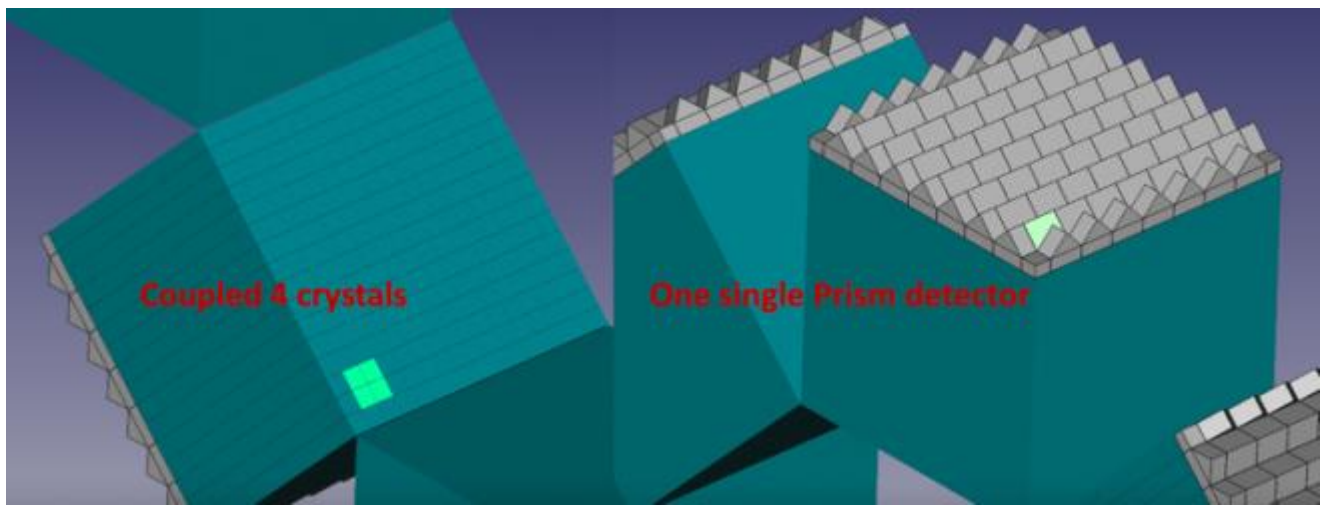

Figure 7: Demonstration of coupled 4 to 1 prism PET – single ring in FreeCAD/GDML WB.

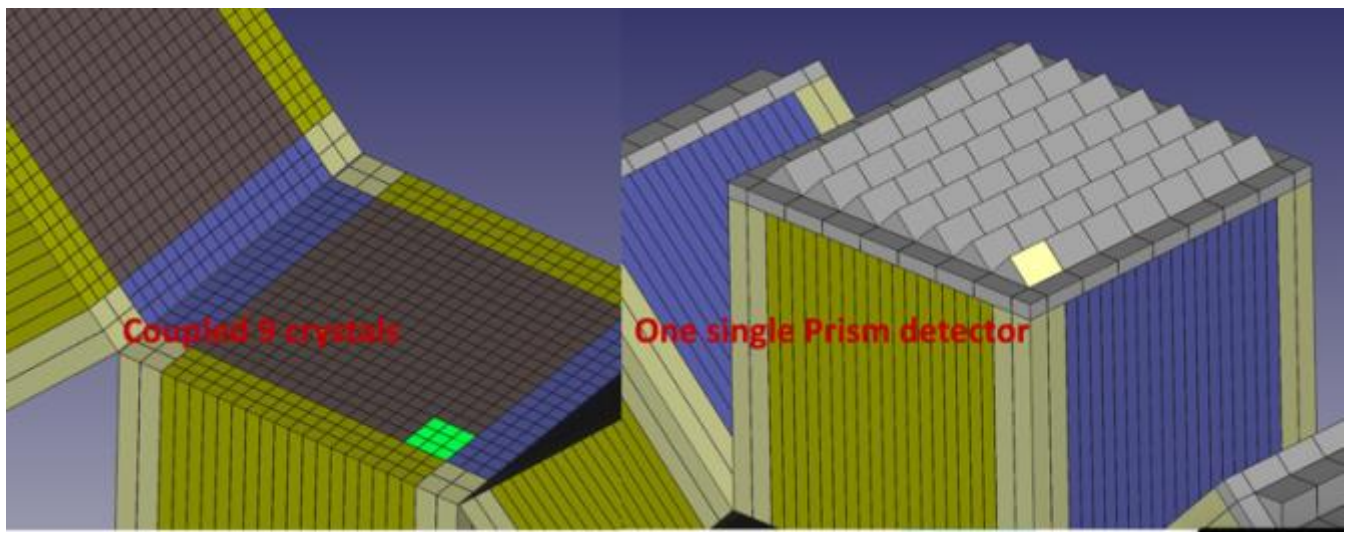

Figure 8: Demonstration of coupled 9 to 1 prism PET – single ring in FreeCAD/GDML WB.

## 2.4 VRAIN PET

This section aims to highlight how the FreeCAD/GDML WB tools can be used to design and construct intricate geometries for specific imaging needs. As an example of a complex structure, we created a VRAIN PET structure. VRAIN was introduced as a dedicated PET system optimized for brain imaging [35–38]. Its design features a hemispherical detector arrangement, which enhances spatial resolution and balances production cost with sensitivity. The system uses 54 detectors in a hemispherical configuration, along with an additional half-ring behind the neck. Each detector consists of a 12 × 12 array of lutetium fine silicate crystals coupled one-to-one with a 12 × 12 array of silicon photomultipliers. This setup reduces photon non-collinearity effects and increases radiation detection efficiency. VRAIN's advanced design results in high-resolution imaging, particularly beneficial for visualizing small brain nuclei and gray matter structures (see [37]).

The VRAIN PET structure starts with a box shape measuring at 4.1 x 4.1 x 10 $mm^3$. This basic element is then replicated into a 12x12 array. Furthermore, the VRAIN PET structure includes variation in the number of elements and the rotational angles in the polar arrays. For instance, creating polar arrays containing 4, 10, 14, and 16 elements, each rotated at specific angles about the x-axis, builds a detailed PET structure. Figure 9 shows the flexibility of polar arrays, where the number of detectors is adjusted based on a ring and the angle of rotation of that ring.

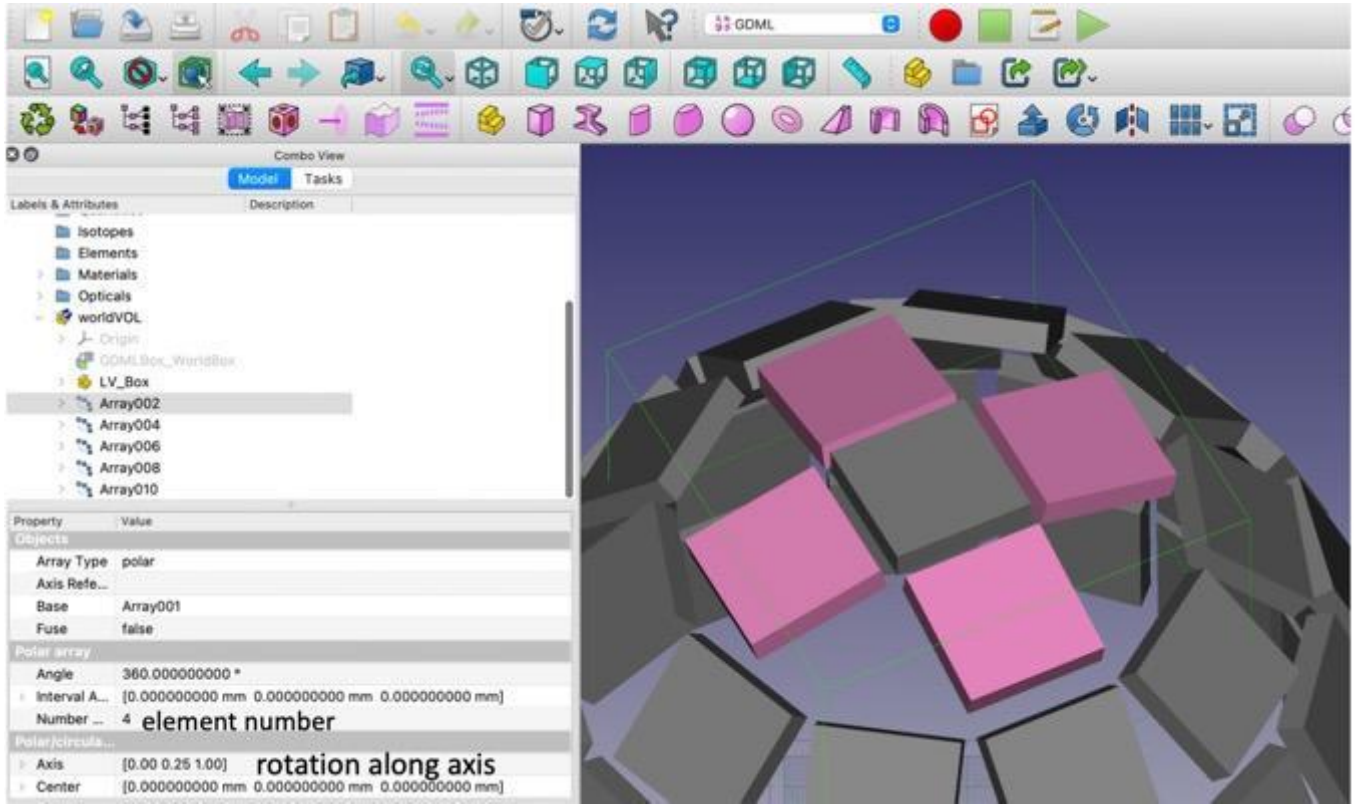

Figure 9: Adjustable polar array tool.

Each ring is then positioned and oriented to form a cohesive whole. The process involves delicate placement and rotation, ensuring that each component aligns correctly within the overall design. Figure 10 shows an overview of the VRAIN PET structure, which demonstrates the flexibility of FreeCAD/GDML to create complex structures [39].

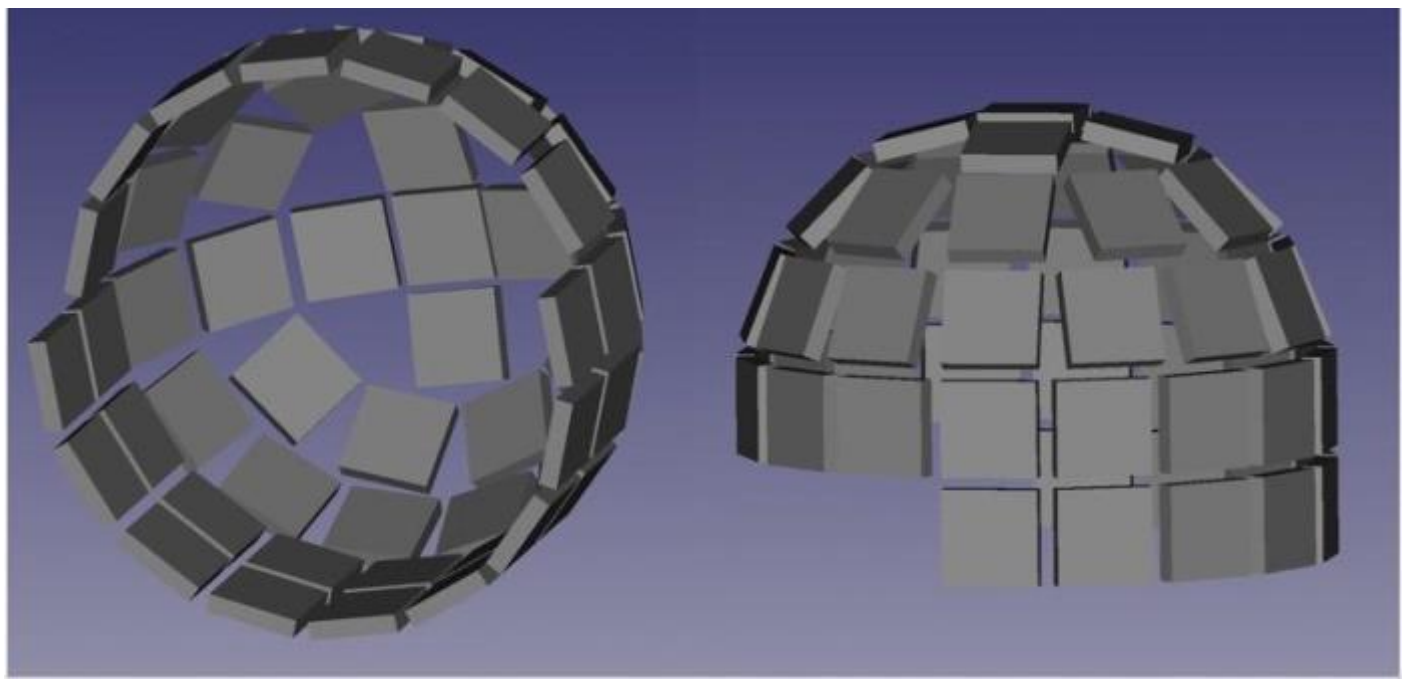

Figure 10: Demonstration of VRAIN PET structure in two views in FreeCAD/GDML.

## 2.5 BRAIN PET

This section demonstrates the automatic process of creating complex objects by using external and macro files. This automation not only streamlines the design process but also ensures precision and consistency in the production of components.

The FreeCAD and GDML WB tools do not support the arbitrary placement of spherical arrays, i.e., arrays in which objects are placed on a spherical surface and point towards the origin. (The supported polar array places objects uniformly around an axis.) To support generation of spherical arrays, we developed a macro file (a Python script), which is customized for placing the objects in arbitrary positions beyond the planar and polar arrangements.

We start by creating a simple box, which serves as the basic unit of the structure. By copying and pasting predefined rotation angles into the Python console, users can manipulate the geometric configurations seamlessly. The integration of these rotation angles allows for precise positioning and orientation of each object within the PET structure.

By selecting the box and executing specific Python commands (see Figure 11), users can define the object locations and orientations through external files containing center points and angles. The same approach can be applied to create the rest of the rings and panels, with adjustments made to correspond to the specific components being designed. By the end, a fully automated process for constructing the PET scanner elements has been developed, with all central objects accurately positioned and oriented. Figure 11 shows how to insert a script into Python the console.

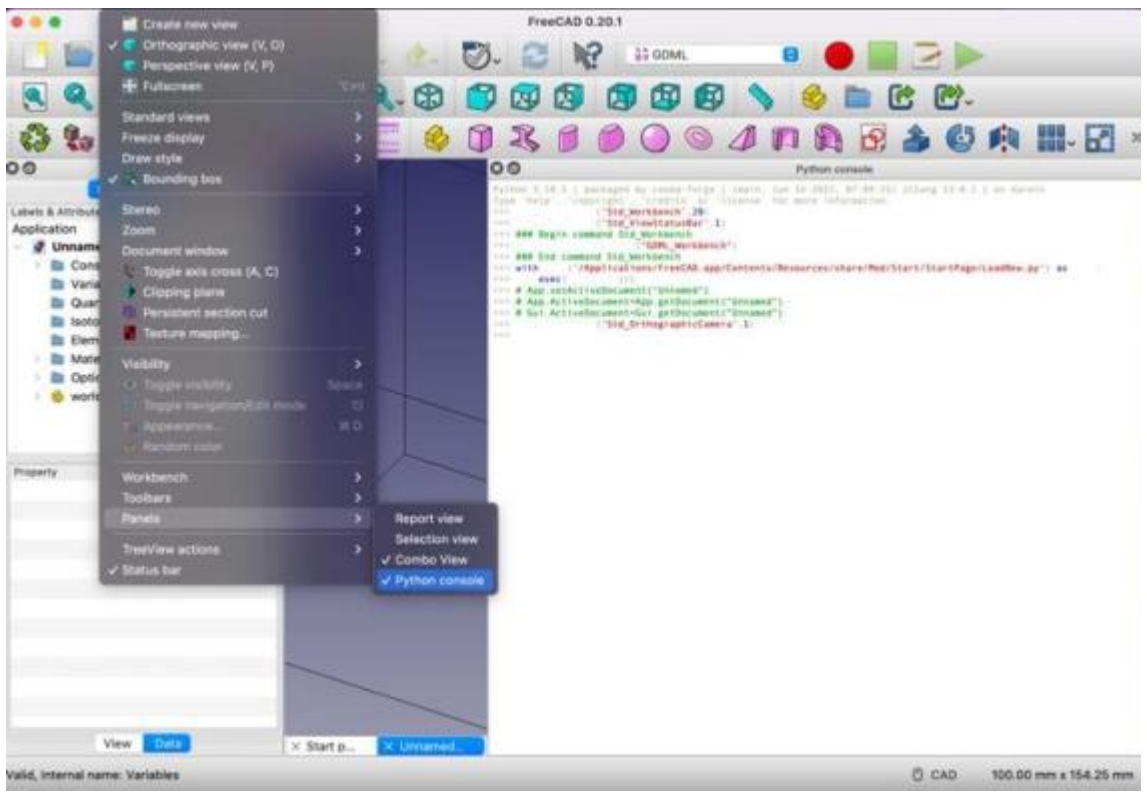

**Figure 11**: Python console allow users integrating script.

The Python script makes the construction of complex PET designs both accessible and efficient. Thus, if users can script a macro file and an external file containing the center points of the desired objects, then these files can be imported into FreeCAD/GDML to automate the generation of specific structures or objects. Figure 12 shows a Brain PET structure created by using an external file and a macro file to place the components of complex PET design (see tutorial video at [40]).

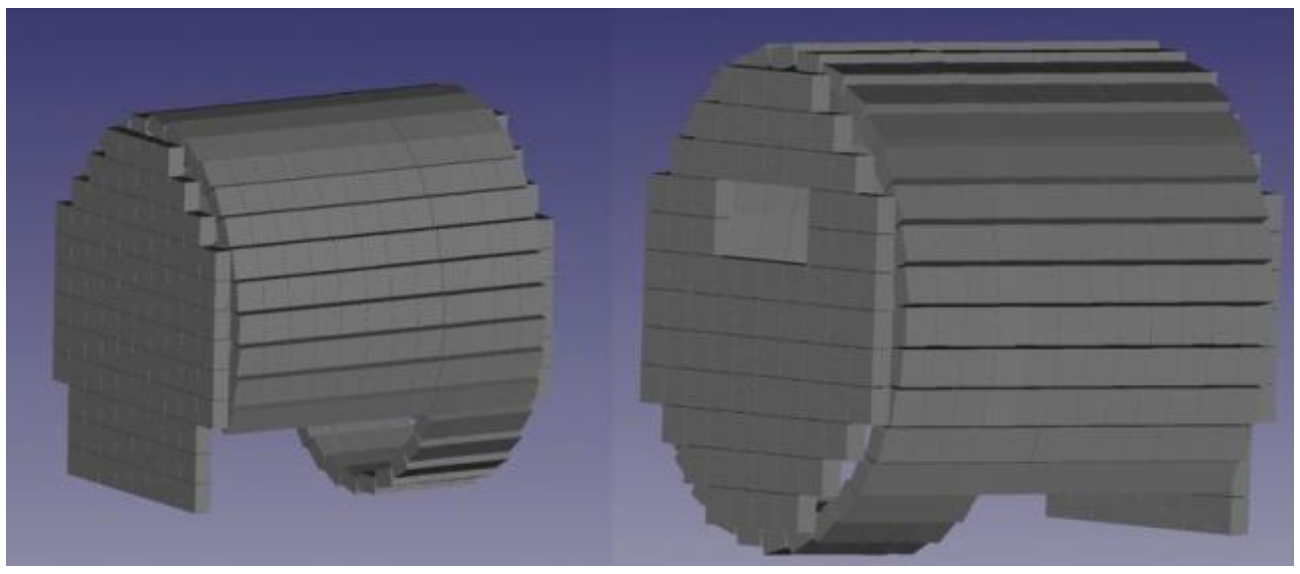

**Figure 12**: Demonstration of BRAIN PET structure in two views in FreeCAD/GDML.

## 3 SINGLE PHOTON EMISSION COMPUTED TOMOGRAPHY (SPECT)

This section shows the development of structural components for common SPECT scanners by showcasing the Dynamic Cardiac-dedicated SPECT scanner developed by our team [41–44]. SPECT scanners comprise gamma cameras, collimators, and detector arrays. Gamma cameras are equipped with scintillation crystals and photomultiplier tubes, converting gamma rays into light and then into electrical signals. Collimators, typically made from lead or tungsten, filter incoming gamma rays to ensure only those at specific angles reach the detectors, thereby enhancing spatial resolution. Detector arrays, arranged in square or multi-headed configurations, ensure comprehensive coverage and high sensitivity [41–44]. These components must be precisely aligned and synchronized to generate precise and high-resolution tomographic images.

Our DC-SPECT scanner comprises fundamental components such as gamma cameras, collimators, and detector arrays. It is important to note that these methodologies and designs are applicable and transferable to other SPECT scanners, demonstrating GDML's ability to provide the necessary design tools across different SPECT scanners.

### 3.1 Detector Design in Arbitrary Array Arrangement

Standard detector configuration begins with the creation of a 25 by 25 array of crystal detectors, each measuring 10 x 2 x 2 $mm^3$, which is replicated to form a detector array that fits design requirements by using a macro file to automate the positioning and orientation of detector boxes with arbitrary placements. The macro file enables users to define coordinates and rotational angles for each detector unit via external data files. Also, it allows the central reference array to be hidden and the manual addition of comments in the GDML file upon export. Figure 13 shows the detector arrays that are created in an automatic manner using external and macro filess in FreeCAD/GDML for the DC-SPECT scanner [45].

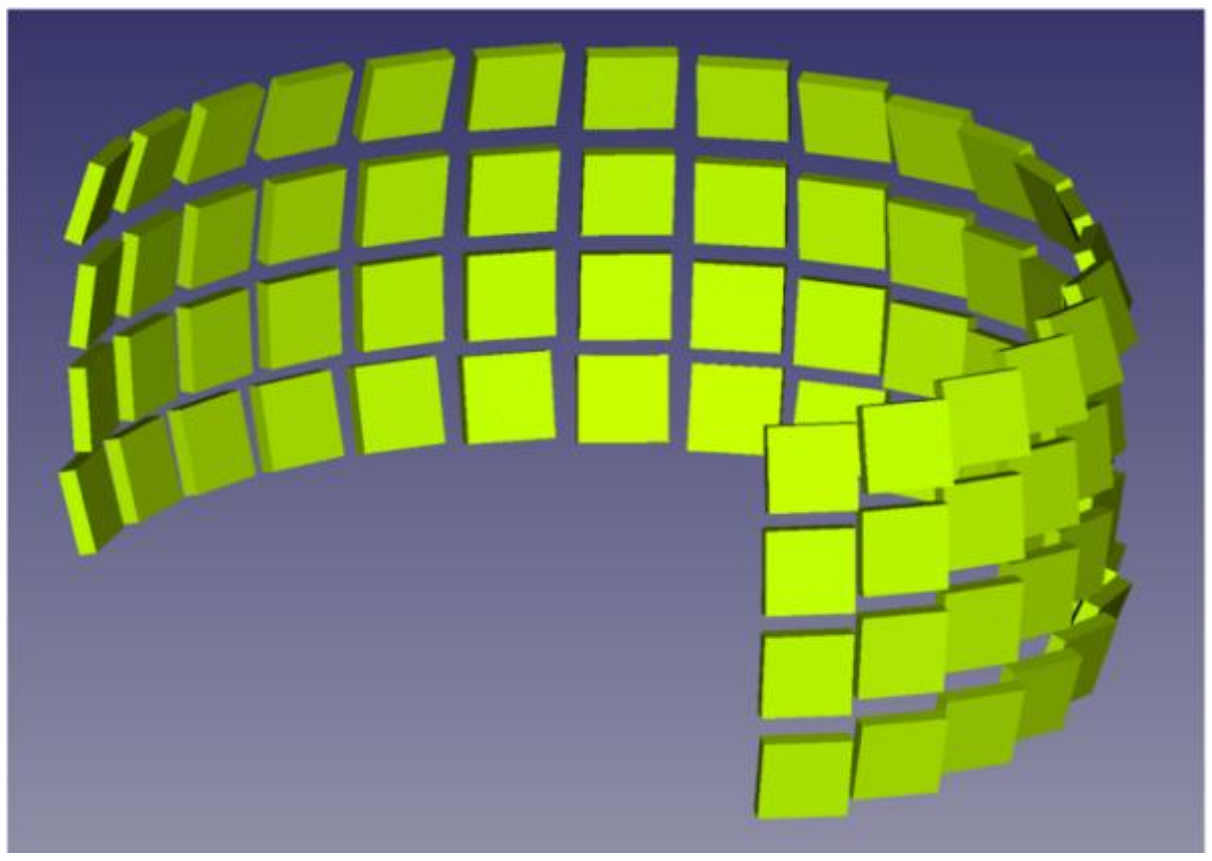

**Figure 13**: Demonstration of detector array structure in FreeCAD/GDML.

### 3.2 Collimator Design

Next, we explore two variations for generating the collimator in FreeCAD/GDML WB. By showing different variations, we demonstrate the possibilities that accommodate the design requirements. Creating multiple variations allows us to evaluate the efficacy of different collimator design in terms of computational simulation and preparation work.

*Variation 1*: This version of the collimator is designed by creating the body part, typically a frustum, and proceeds by constructing the base and pinhole. Each component is designed using the FreeCAD/GDML WB, with precise dimensions and spatial relations to ensure proper alignment. For example, a square frustum may be created with dimensions of x1 = x2 = y1 = 48 mm, x3 = x4 = y2 = 6.3 mm, and z = 78 mm. This frustum is then refined by subtracting a smaller frustum from it, resulting in a thin-walled, truncated pyramidal structure. Similarly, the base and pinhole are constructed by creating boxes and smaller frustums and refining their dimensions through a series of cuts, ensuring they align perfectly with the pyramidal body. Figure 14 shows a DC-SPECT collimator created following the first variation [46].

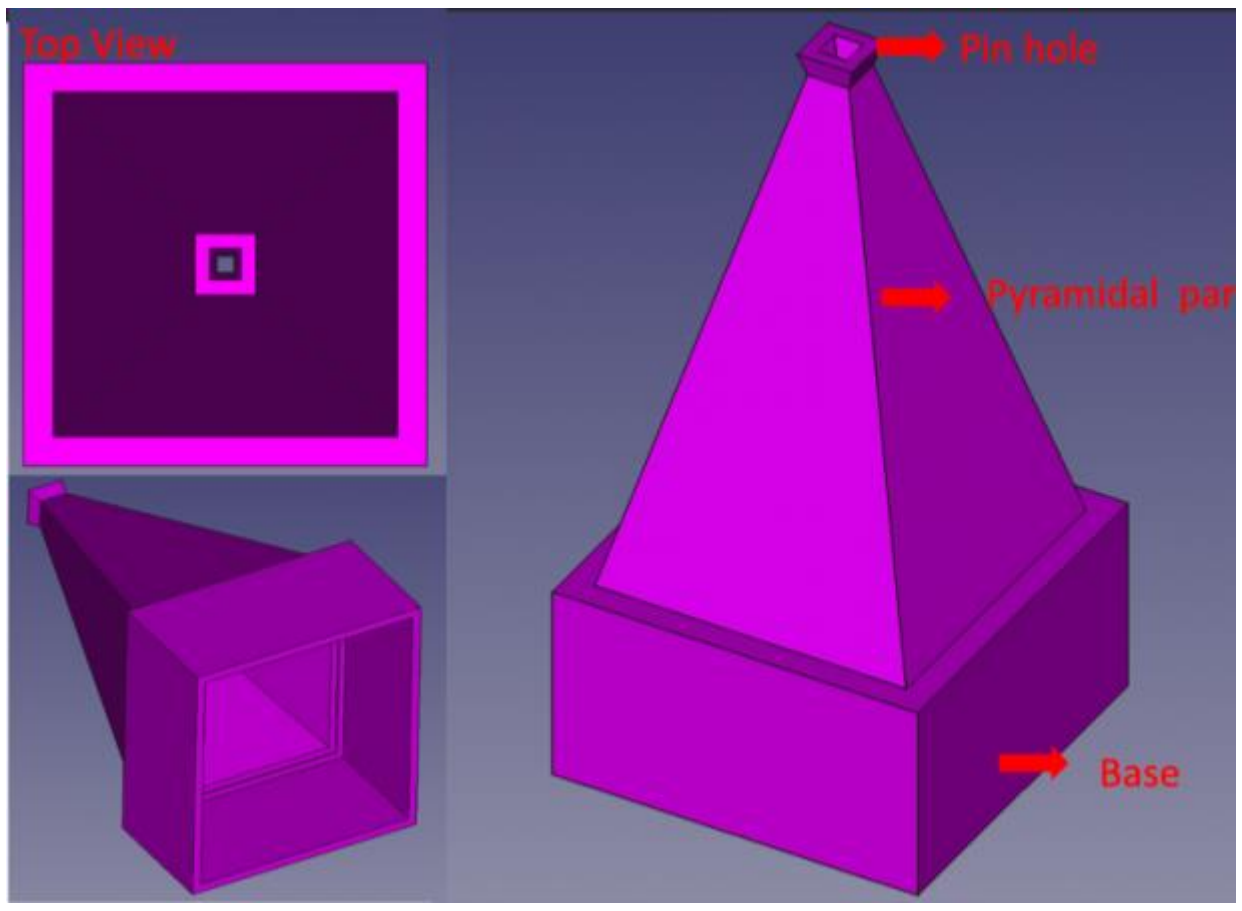

**Figure 14**: Demonstration of variation 1 - collimator structure in three views in FreeCAD/GDML.

*Variation 2*: Another method for constructing the collimator model involves merging pyramidal, pinhole, and box walls. In this approach, we use the "trap" (trapezoid) tool to create four trapezoidal walls, adjusting their placements and rotations to form a thin pyramidal body. The base is similarly constructed by creating and positioning box walls, while the pinhole is built from smaller box walls and positioned precisely. This variation shows an alternative way to create collimators without using the cut tool. Figure 15 illustrates a DC-SPECT collimator created using this second variation.

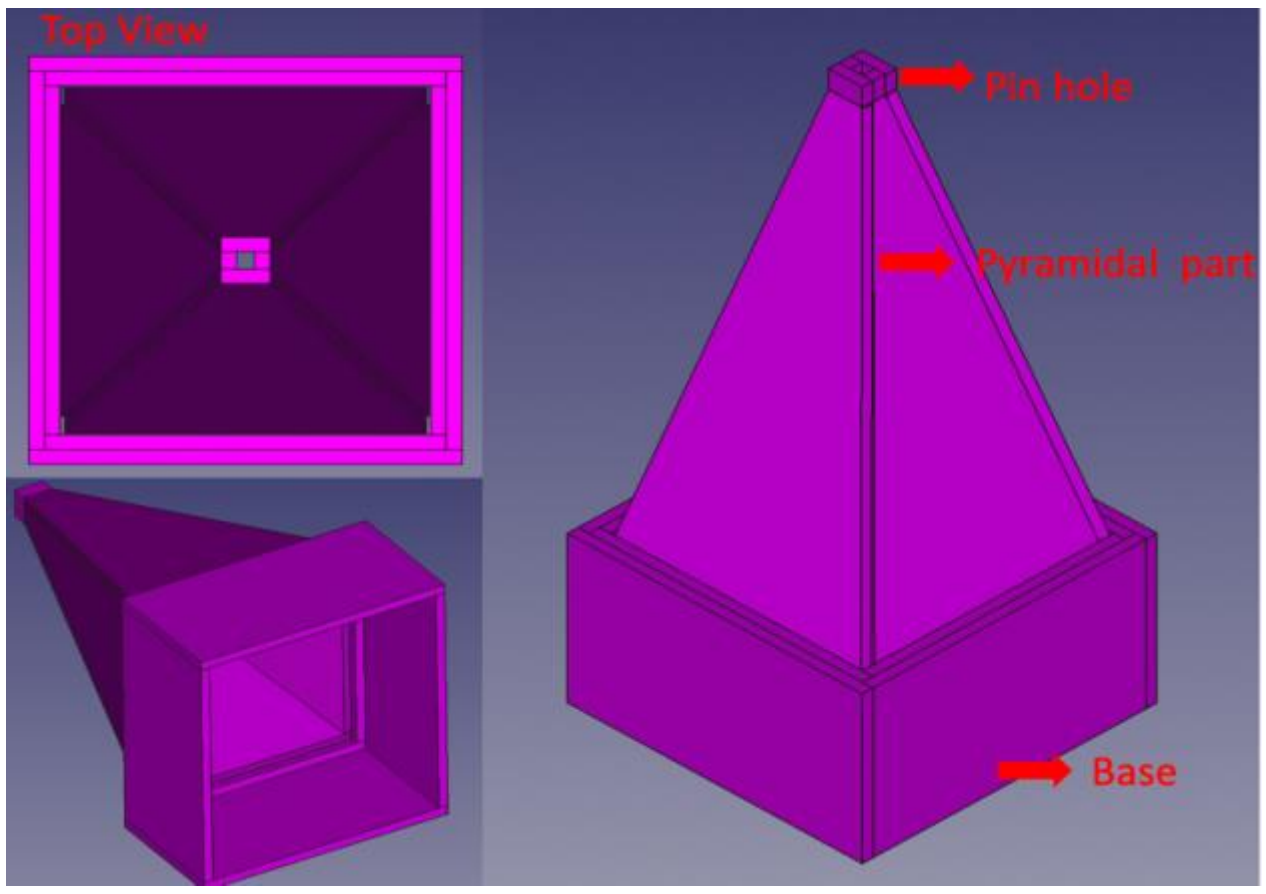

**Figure 15**: Demonstration of variation 2 - collimator structure in three views in FreeCAD/GDML.

The “union” tool can be used to merge all three parts into a single collimator, or they can be grouped into the same box/tree for easier positioning and orientation as a whole unit. Depending on the design requirements and performance efficiency, users can choose to merge only two parts, such as the pyramidal body and pinhole, and separate the base part in another box/tree. One of the advantages of FreeCAD/GDML is the ability to record macros, allowing users to record the collimator creation process with varying dimensions. By adjusting the values in the macro file, the tool can automatically generate different collimator configurations. Figure 16 shows collimator arrays that are automatically created and placed using a recorded macro file.

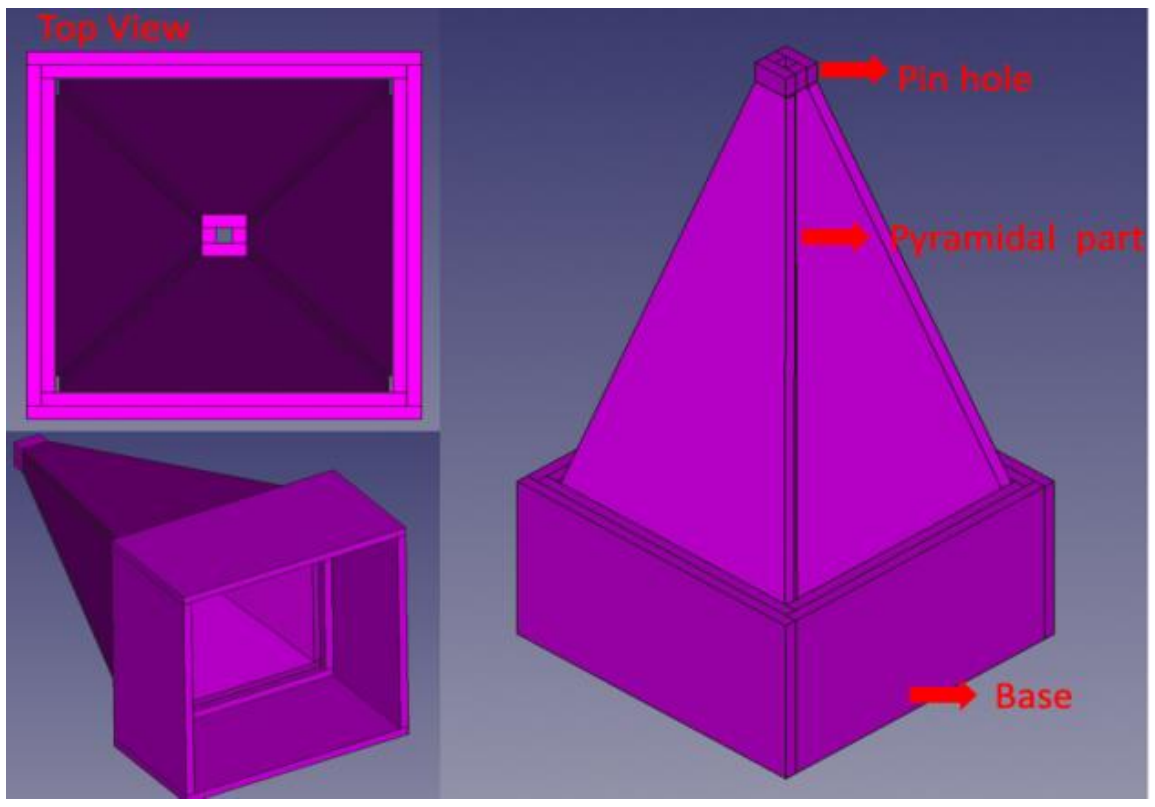

**Figure 16**: Demonstration of collimator array structure in FreeCAD/GDML.

## 3.3 Combining Variations for a Complete SPECT Scanner

The combination of various collimator designs and detector configurations yields a complete SPECT scanner. Additionally, the ability to import components such as lead shields further enhances the scanner's functionality. These shields reduce background noise and improve image quality by blocking unwanted radiation. This modular approach allows for a customizable SPECT scanner design, tailored to meet the needs of scientific research and medical diagnostics. Figures 17 & 18 show the full array of DC-SPECT gamma cameras, including the collimator shells and the scintillator crystals, without and with the shield, respectively, that we automatically create using external and macro files [47].

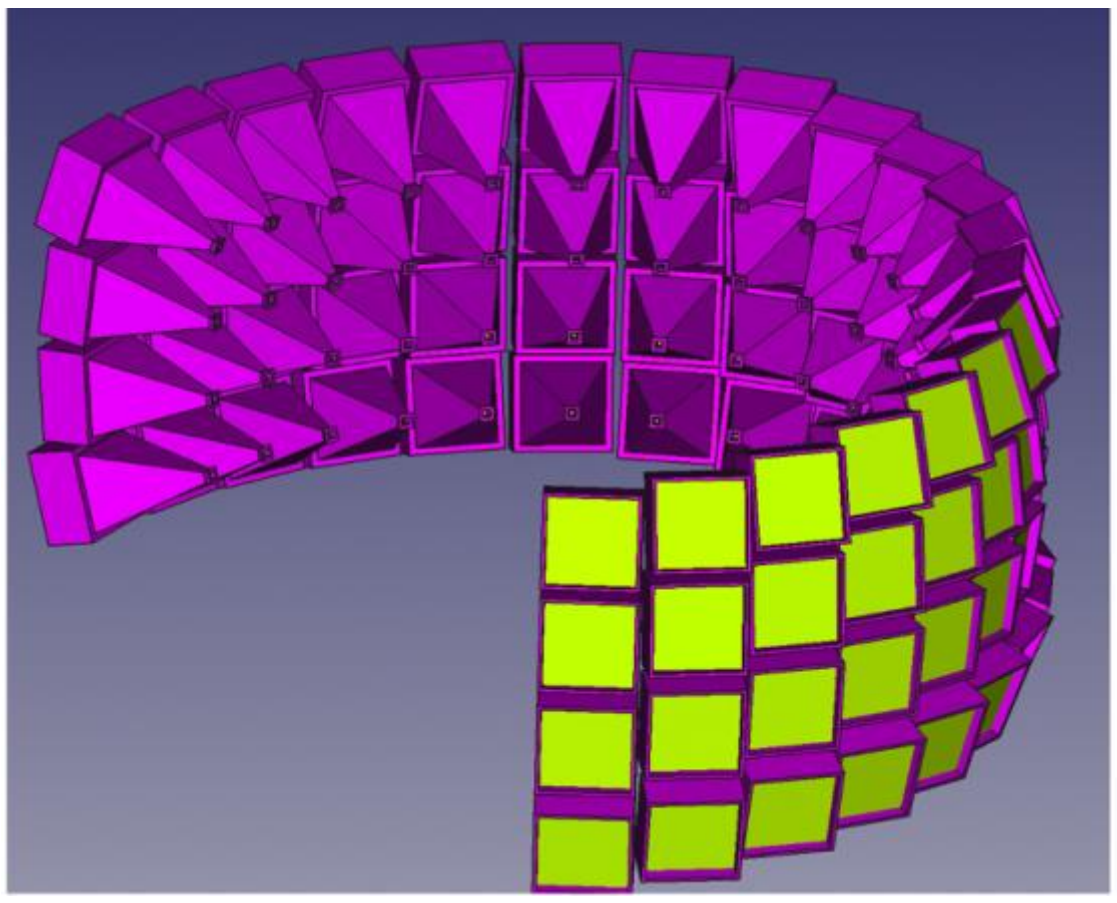

**Figure 17**: Demonstration of DC-SPECT structure in FreeCAD/GDML.

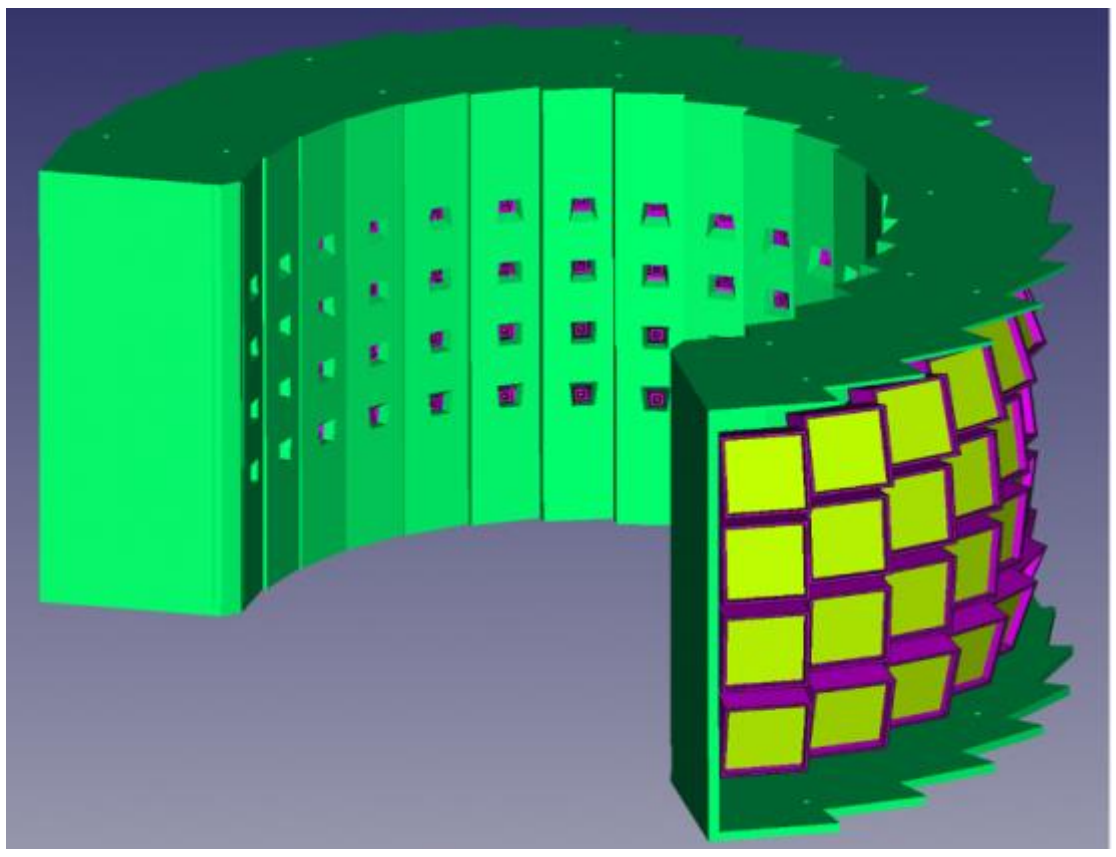

Figure 18: Demonstration of DC-SPECT with shield in FreeCAD/GDML.

## 4 INTEGRATION OF SPECT AND PET IN A SINGLE STRUCTURE

The integration of SPECT and PET into a single structure represents a potential advancement in medical imaging technology. This integration facilitates simultaneous acquisition of functional and metabolic information, improving diagnostic accuracy and offering a comprehensive view of physiological processes [48,49]. Within FreeCAD/GDML, users can design and assemble intricate structures combining SPECT and PET components. Leveraging geometric modeling and scripting capabilities, various detector configurations and collimator designs can be seamlessly integrated into unified systems. Additionally, the flexibility of GDML permits the inclusion of supplementary components, such as lead shields, to improve image quality and reduce stray radiation exposure.

## 5 INTEGRATION OF SPECT AND PET IN A SINGLE STRUCTURE

Phantoms are essential tools in medical imaging and radiation therapy, serving as stand-ins for human tissues or other materials to test, calibrate, and validate imaging devices and techniques [50–53]. These structures mimic the physical and geometric properties of the tissues they represent, enabling precise experimentation and optimization of imaging protocols without the need for human subjects. In this context, we introduce a process to create phantoms that exemplifies the use of triangular arrays within the GDML WB to automate the design of patterned structures. We developed a script for "Triangular Array", which can be accessed from the FreeCAD Python console. This macro enables users to select and replicate an object. Figure 19 shows an overview of a triangular array Phantom [54].

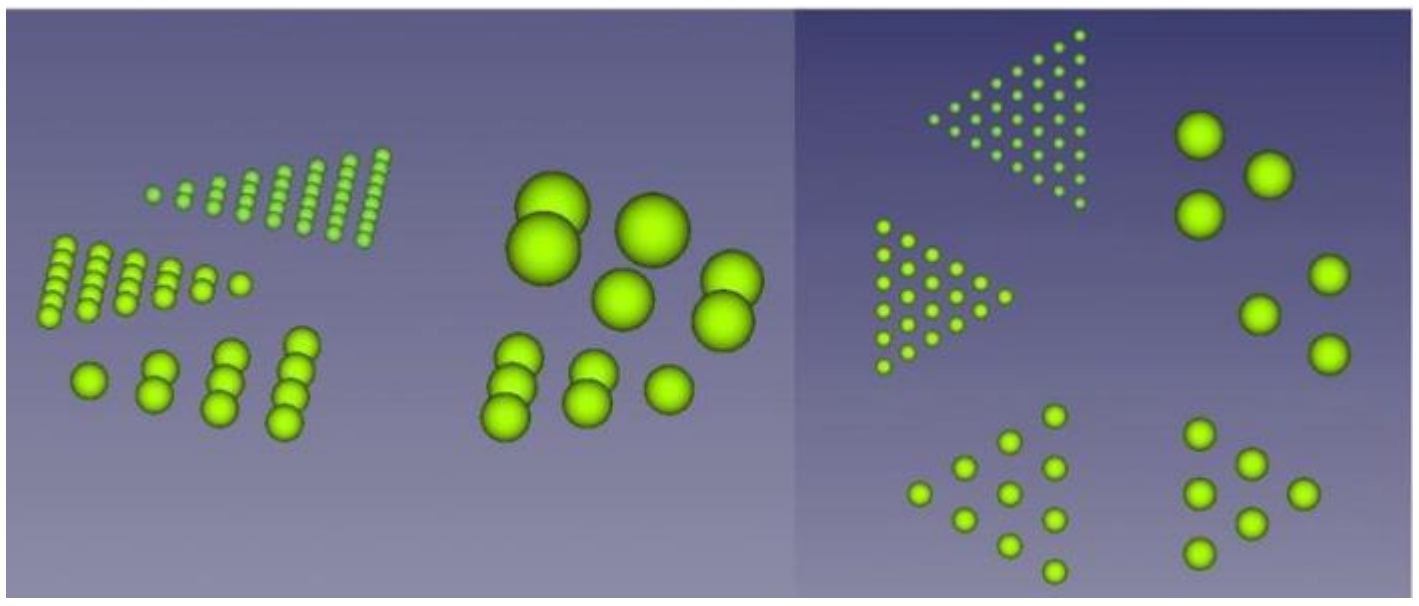

Figure 19: Demonstration of phantom in two views in FreeCAD/GDML.

# 6 OTHER FEATURES OF THE FREECAD/GDML WB

## 6.1 Converting Structures from External CAD Platforms into the FreeCAD/GDML WB

Converting structures from other CAD platforms (such as SolidWorks, Fusion, OnShape and others) into FreeCAD/GDML involves: (1) exporting a model as a STEP file from the CAD package and ensuring the tessellation parameters capture all details; and (2) importing the STEP file into FreeCAD by using the File -> Import -> "step file" menu item. This process transforms the model into a tessellated mesh. Tessellated meshing in the GDML Workbench is crucial for maintaining geometric fidelity necessary for accurate simulations, particularly in applications requiring intricate surface details. This conversion process enables us to leverage the detailed designs created outside of FreeCAD for advanced simulations on GDML-compatible platforms. Figure 20 demonstrates how to import a step file into the FreeCAD/GDML workbench.

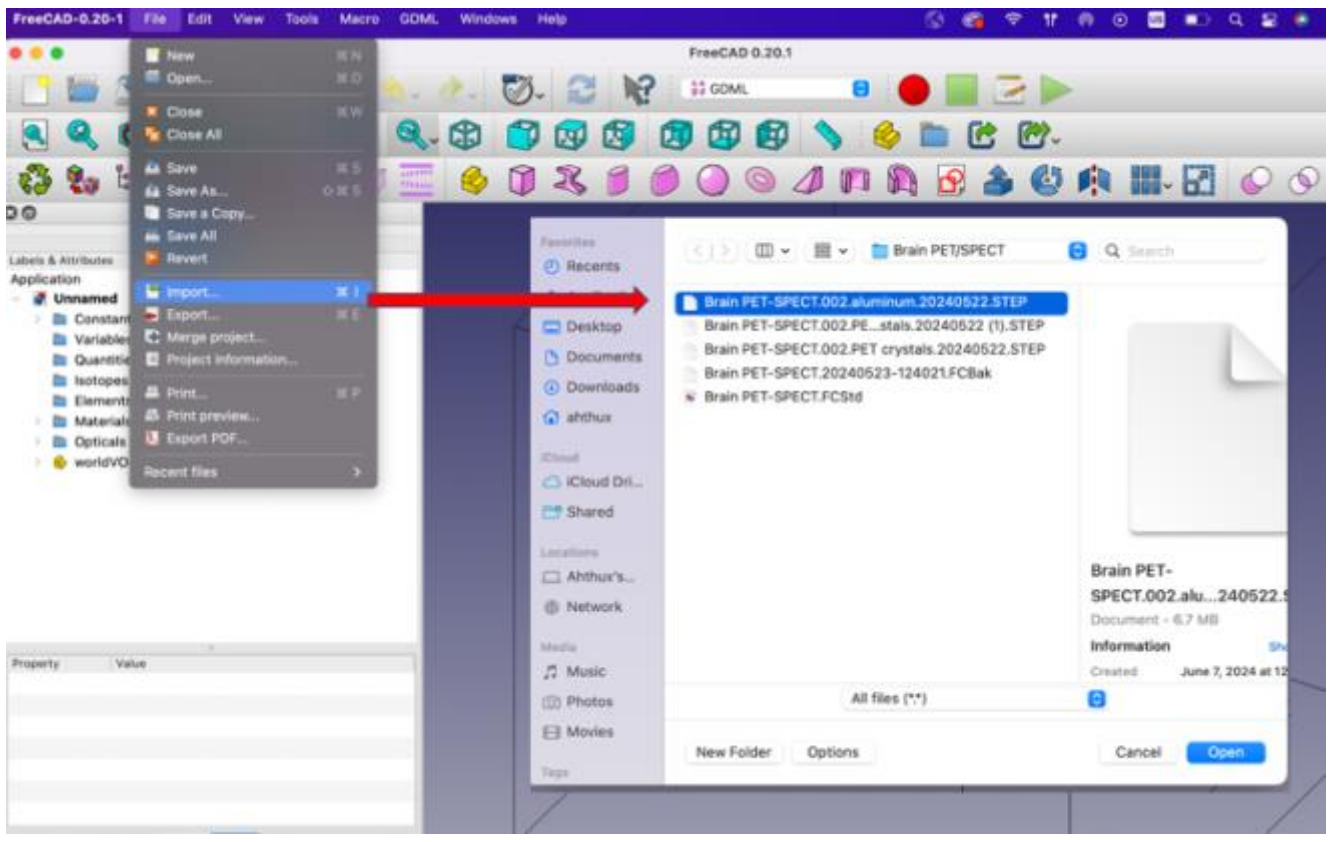

**Figure 20**: Demonstration of Import STEP file into FreeCAD/GDML WB.

## 6.2 Importing Optical/Material and Exporting to GDML

Integrating optical and material properties into GDML files in FreeCAD enables the external assignment of these properties to created objects. When FreeCAD/GDML WB lacks specific materials, users can import custom properties from external files, such as XML files. This imported data integrates seamlessly into the GDML WB framework, expanding the utility of simulations by accommodating a broader range of materials. The material format follows the Geant4 protocol and is usable in simulations. Figure 21 illustrates how to import optical/material files into FreeCAD/GDML workbench [55], while Figure 22 shows how to export a GDML file from the FreeCAD/GDML workbench.

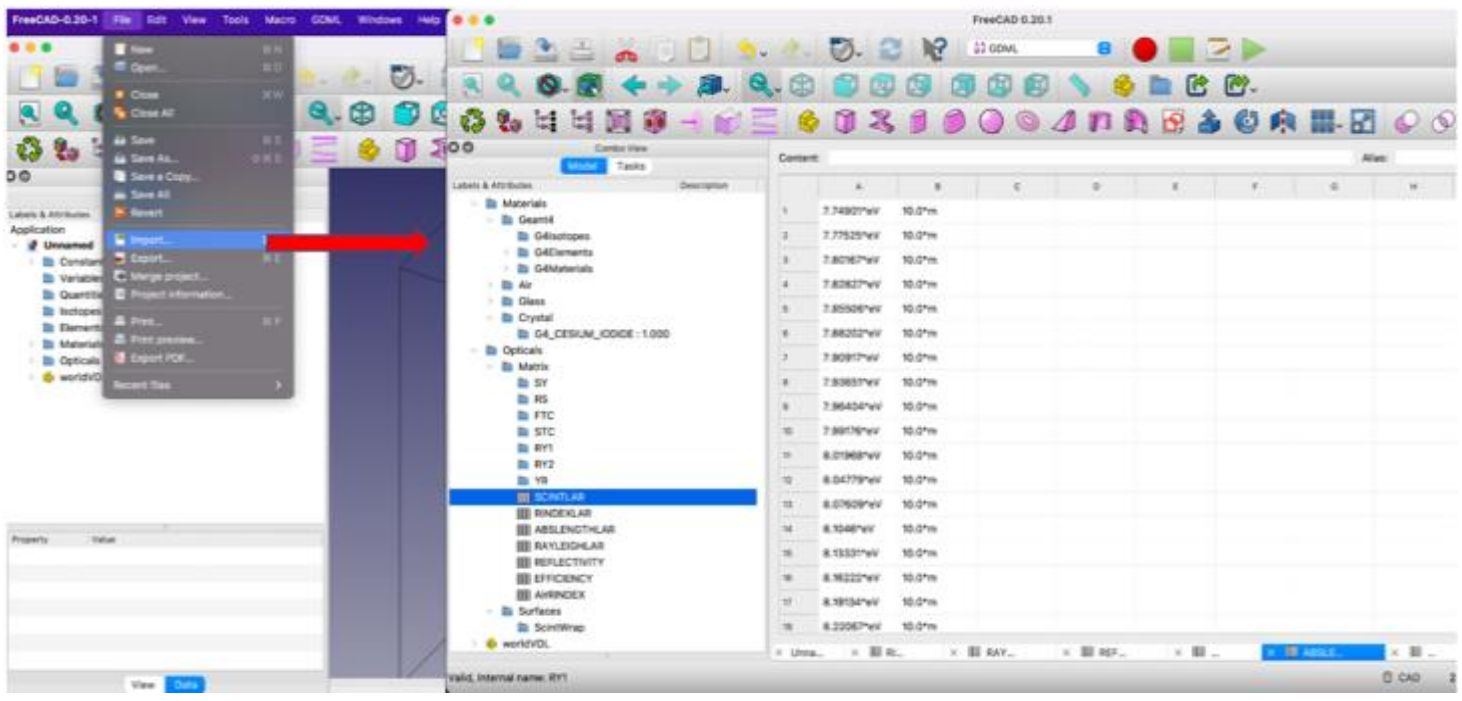

**Figure 21**: Demonstration of Import optical/materials into FreeCAD/GDML WB.

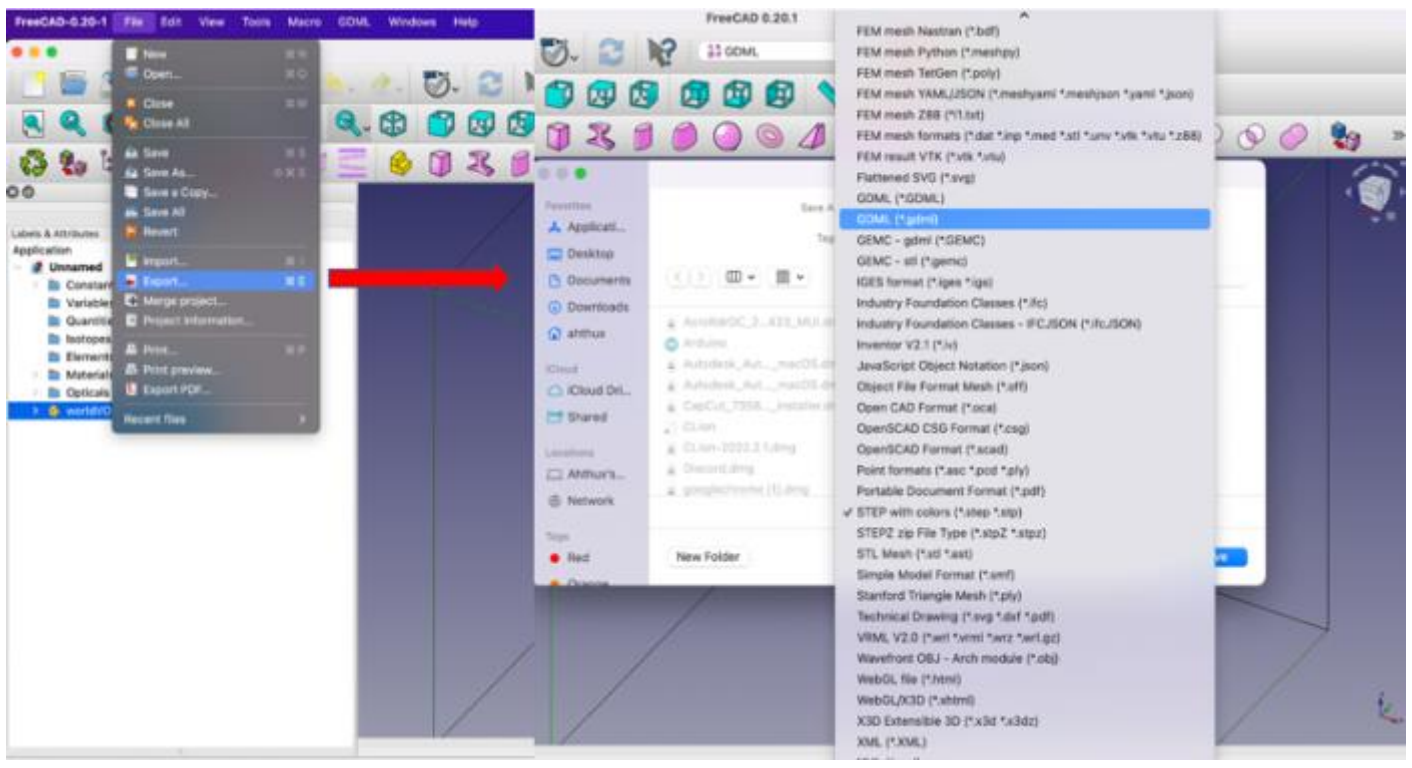

**Figure 22**: Demonstration of Export as GDML file from FreeCAD/GDML WB.

## 7 CONCLUSION

This work demonstrates the flexibility of FreeCAD/GDML in designing various medical imaging applications. These tools offer robust capabilities for geometric modeling, scripting, and integration, allowing for precise customization of key scanner components, including detectors, collimators, and structural assemblies. This adaptability allows researchers and engineers to tailor scanner designs to meet the specific demands of advanced imaging modalities, especially in optimizing performance and improving diagnostic accuracy in PET and SPECT imaging. FreeCAD/GDML has already been successfully implemented in high-performance simulation platforms like VPG4 [13], and we anticipate its adoption will become increasingly widespread within the broader medical imaging community.

## ACKNOWLEDGMENTS

We sincerely thank Keith Sloan and Munther Hindi for their help with FreeCAD and GDML, which was essential to this work and final edit of this manuscript. We acknowledge support from NSF grant: med240002p, enabling computational simulations through the ACCESS program clusters. This work is partially supported by the NIH grant R01HL145160.

## REFERENCES

[1] Roncali E and Cherry S R 2011 Application of Silicon Photomultipliers to Positron Emission Tomography *Ann. Biomed. Eng.* **39** 1358–77

[2] Di Carli M F, Dorbala S, Meserve J, El Fakhri G, Sitek A and Moore S C 2007 Clinical Myocardial Perfusion PET/CT *J. Nucl. Med.* **48** 783–93

[3] Bar-Shalom R, Yefremov N, Guralnik L, Gaitini D, Frenkel A, Kuten A, Altman H, Keidar Z and Israel O 2003 Clinical performance of PET/CT in evaluation of cancer: additional value for diagnostic imaging and patient management *J. Nucl. Med. Off. Publ. Soc. Nucl. Med.* **44** 1200–9

[4] Liu Z, Niu M, Kuang Z, Ren N, Wu S, Cong L, Wang X, Sang Z, Williams C and Yang Y 2022 High resolution detectors for whole-body PET scanners by using dual-ended readout *EJNMMI Phys.* **9** 29

[5] Bondarenko G, Dolgoshein B, Golovin V, Ilyin A, Klanner R and Popova E 1998 Limited Geiger-mode silicon photodiode with very high gain *Nucl. Phys. B - Proc. Suppl.* **61** 347–52
[6] Spanoudaki V Ch and Levin C S 2010 Photo-Detectors for Time of Flight Positron Emission Tomography (ToF-PET) *Sensors* **10** 10484–505
[7] Moses W W 2011 Fundamental limits of spatial resolution in PET *Nucl. Instrum. Methods Phys. Res. Sect. Accel. Spectrometers Detect. Assoc. Equip.* **648** S236–40
[8] Schaart D R 2021 Physics and technology of time-of-flight PET detectors *Phys. Med. Biol.* **66** 09TR01
[9] Gross I, Rooney S A and Warshaw J B 1975 The influence of cortisol on the enzymes of fatty acid synthesis in developing mammalian lung and brain *Pediatr. Res.* **9** 752–5
[10] Wikipedia contributors 2024 FreeCAD *Wikipedia Free Encycl.*
[11] Sloan K 2024 FreeCAD GDML Workbench - AddonManager Installable
[12] Agostinelli S, Allison J, Amako K, Apostolakis J, Araujo H, Arce P, Asai M, Axen D, Banerjee S, Barrand G, Behner F, Bellagamba L, Boudreau J, Broglia L, Brunengo A, Burkhardt H, Chauvie S, Chuma J, Chytracek R, Cooperman G, Cosmo G, Degtyarenko P, Dell'Acqua A, Depaola G, Dietrich D, Enami R, Feliciello A, Ferguson C, Fesefeldt H, Folger G, Foppiano F, Forti A, Garelli S, Giani S, Giannitrapani R, Gibin D, Gómez Cadenas J J, González I, Gracia Abril G, Greeniaus G, Greiner W, Grichine V, Grossheim A, Guatelli S, Gumplinger P, Hamatsu R, Hashimoto K, Hasui H, Heikkinen A, Howard A, Ivanchenko V, Johnson A, Jones F W, Kallenbach J, Kanaya N, Kawabata M, Kawabata Y, Kawaguti M, Kelner S, Kent P, Kimura A, Kodama T, Kokoulin R, Kossov M, Kurashige H, Lamanna E, Lampén T, Lara V, Lefebure V, Lei F, Liendl M, Lockman W, Longo F, Magni S, Maire M, Medernach E, Minamimoto K, Mora De Freitas P, Morita Y, Murakami K, Nagamatu M, Nartallo R, Nieminen P, Nishimura T, Ohtsubo K, Okamura M, O'Neale S, Oohata Y, Paech K, Perl J, Pfeiffer A, Pia M G, Ranjard F, Rybin A, Sadilov S, Di Salvo E, Santin G, Sasaki T, et al 2003 Geant4—a simulation toolkit *Nucl. Instrum. Methods Phys. Res. Sect. Accel. Spectrometers Detect. Assoc. Equip.* **506** 250–303
[13] Hashemi A, Ottensmeyer M, Feng Y and Sabet H 2023 Versatile Geant4 simulation application with high performance for complex nuclear medicine imaging scanners *Soc. Nucl. Med.*
[14] Hashemi A, Ottensmeyer M, Feng Y and Sabet H 2023 VPG4, Versatile Parallelizable Geant4 Interface: A Novel Platform for Modeling Complex Nuclear Medicine Imaging Scanners
[15] Levin C S and Hoffman E J 1999 Calculation of positron range and its effect on the fundamental limit of positron emission tomography system spatial resolution *Phys. Med. Biol.* **44** 781–99
[16] Gonzalez-Montoro A, Pourashraf S, Cates J W and Levin C S 2022 Cherenkov Radiation–Based Coincidence Time Resolution Measurements in BGO Scintillators *Front. Phys.* **10** 816384
[17] Wagatsuma K, Miwa K, Sakata M, Oda K, Ono H, Kameyama M, Toyohara J and Ishii K 2017 Comparison between new-generation SiPM-based and conventional PMT-based TOF-PET/CT *Phys. Med.* **42** 203–10
[18] Phelps M E 2000 Positron emission tomography provides molecular imaging of biological processes *Proc. Natl. Acad. Sci.* **97** 9226–33
[19] *Simple PET structure- single ring* YouTube uploaded by Anh Le in June 2024 https://www.youtube.com/watch?v=i5byzeuFKgw
[20] Molinos C, Sasser T, Salmon P, Gsell W, Viertl D, Massey J C, Mińczuk K, Li J, Kundu B K, Berr S, Correcher C, Bahadur A, Attarwala A A, Stark S, Junge S, Himmelreich U, Prior J O, Laperre K, Van Wyk S and Heidenreich M 2019 Low-Dose Imaging in a New Preclinical Total-Body PET/CT Scanner *Front. Med.* **6** 88
[21] Moskal P and Stępień E Ł 2020 Prospects and Clinical Perspectives of Total-Body PET Imaging Using Plastic Scintillators *PET Clin.* **15** 439–52
[22] Alavi A, Werner T J, Stępień E Ł and Moskal P 2022 Unparalleled and revolutionary impact of PET imaging on research and day to day practice of medicine *Bio-Algorithms Med-Syst.* **17** 203–12
[23] *Whole body PET - single ring* YouTube uploaded by Anh Le in June 2024 https://www.youtube.com/watch?v=jP5MKFqQtX8

[24] Miyaoka R S, Lewellen T K, Yu H and McDaniel D L 1998 Design of a depth of interaction (DOI) PET detector module *IEEE Trans. Nucl. Sci.* **45** 1069–73

[25] LaBella A, Cao X, Petersen E, Lubinsky R, Biegon A, Zhao W and Goldan A H 2020 High-Resolution Depth-Encoding PET Detector Module with Prismatoid Light-Guide Array *J. Nucl. Med.* **61** 1528–33

[26] Pizzichemi M, Polesel A, Stringhini G, Gundacker S, Lecoq P, Tavernier S, Paganoni M and Auffray E 2019 On light sharing TOF-PET modules with depth of interaction and 157 ps FWHM coincidence time resolution *Phys. Med. Biol.* **64** 155008

[27] Slifstein M and Abi-Dargham A 2017 Recent Developments in Molecular Brain Imaging of Neuropsychiatric Disorders *Semin. Nucl. Med.* **47** 54–63

[28] Surti S 2015 Update on Time-of-Flight PET Imaging *J. Nucl. Med.* **56** 98–105

[29] Van Sluis J, De Jong J, Schaar J, Noordzij W, Van Snick P, Dierckx R, Borra R, Willemsen A and Boellaard R 2019 Performance Characteristics of the Digital Biograph Vision PET/CT System *J. Nucl. Med.* **60** 1031–6

[30] Van Sluis J, De Jong J, Schaar J, Noordzij W, Van Snick P, Dierckx R, Borra R, Willemsen A and Boellaard R 2019 Performance Characteristics of the Digital Biograph Vision PET/CT System *J. Nucl. Med.* **60** 1031–6

[31] Stickel J R and Cherry S R 2005 High-resolution PET detector design: modelling components of intrinsic spatial resolution *Phys. Med. Biol.* **50** 179–95

[32] LaBella A, Cao X, Petersen E, Lubinsky R, Biegon A, Zhao W and Goldan A H 2020 High-Resolution Depth-Encoding PET Detector Module with Prismatoid Light-Guide Array *J. Nucl. Med.* **61** 1528–33

[33] *Prism PET - Coupled 4 to 1* YouTube uploaded by Anh Le in June 2024 https://www.youtube.com/watch?v=aJVQkEk7oTE

[34] *Prism PET - Coupled 9 to 1* YouTube uploaded by Anh Le in June 2024 https://www.youtube.com/watch?v=CnkCkF6rWr8

[35] Yang Y, Wu Y, Qi J, St. James S, Du H, Dokhale P A, Shah K S, Farrell R and Cherry S R 2008 A Prototype PET Scanner with DOI-Encoding Detectors *J. Nucl. Med.* **49** 1132–40

[36] Akamatsu G, Tashima H, Yoshida E, Wakizaka H, Iwao Y, Maeda T, Takahashi M and Yamaya T 2019 Modified NEMA NU-2 performance evaluation methods for a brain-dedicated PET system with a hemispherical detector arrangement *Biomed. Phys. Eng. Express* **6** 015012

[37] Akamatsu G, Takahashi M, Tashima H, Iwao Y, Yoshida E, Wakizaka H, Kumagai M, Yamashita T and Yamaya T 2022 Performance evaluation of VRAIN: a brain-dedicated PET with a hemispherical detector arrangement *Phys. Med. Biol.* **67** 225011

[38] Ahmed A M, Tashima H, Yoshida E, Nishikido F and Yamaya T 2017 Simulation study comparing the helmet-chin PET with a cylindrical PET of the same number of detectors *Phys. Med. Biol.* **62** 4541–50

[39] *Vrain PET* YouTube uploaded by Anh Le in June 2024 https://www.youtube.com/watch?v=umb90JBEFo8

[40] *Brain PET* YouTube uploaded by Anh Le in June 2024 https://www.youtube.com/watch?v=Wch8mkxBvwI

[41] Feng Y, Hashemi A, Soleymani S, Ottensmeyer M and Sabet H 2022 DB-SPECT, a Fixed-Gantry SPECT Scanner for Dynamic Brain Imaging: Design Concept and First Results *2022 IEEE Nuclear Science Symposium and Medical Imaging Conference (NSS/MIC)* 2022 IEEE Nuclear Science Symposium and Medical Imaging Conference (NSS/MIC) (Italy: IEEE) pp 1–3

[42] Sabet H, Haoning Liang, Yusheng Li, and Wei Chang 2010 Development of a modular detector system for C-SPECT *IEEE Nuclear Science Symposuim & Medical Imaging Conference* 2010 IEEE Nuclear Science Symposium and Medical Imaging Conference (2010 NSS/MIC) (Knoxville, TN: IEEE) pp 2545–8

[43] Blackberg L, Sajedi S, Anderson O A, Feng Y, Fakhri G E, Furenlid L and Sabet H 2020 Dynamic Cardiac SPECT for diagnostic and theranostics applications: latest results *2020 IEEE Nuclear Science Symposium and Medical Imaging Conference (NSS/MIC)* 2020 IEEE Nuclear Science Symposium and Medical Imaging Conference (NSS/MIC) (Boston, MA, USA: IEEE) pp 1–3

[44] Hamid Sabet, Sajedi S, Blackberg L and Vittum B 2022 Dynamic Cardiac SPECT system for diagnostic and theranostics applications: design optimization and experimental data *J. Nucl. Med.*

[45] *Detector* YouTube uploaded by Anh Le in June 2024 https://www.youtube.com/watch?v=B48fkUH-4Rc

[46] *Collimator_head 1* YouTube uploaded by Anh Le in June 2024 https://www.youtube.com/watch?v=opMh2aRWPBA

[47] *Cardiac SPECT* YouTube uploaded by Anh Le in June 2024 https://www.youtube.com/watch?v=A_8jFCBHXU8

[48] Adams M C, Turkington T G, Wilson J M and Wong T Z 2010 A Systematic Review of the Factors Affecting Accuracy of SUV Measurements *Am. J. Roentgenol.* **195** 310–20

[49] Kinahan P E, Townsend D W, Beyer T and Sashin D 1998 Attenuation correction for a combined 3D PET/CT scanner *Med. Phys.* **25** 2046–53

[50] Bell D and Hacking C 2020 Phantom *Radiopaedia.org* (Radiopaedia.org)

[51] Leng S, Yu L, Vrieze T, Kuhlmann J, Chen B and McCollough C H 2015 Construction of realistic liver phantoms from patient images using 3D printer and its application in CT image quality assessment SPIE Medical Imaging ed C Hoeschen, D Kontos and T G Flohr (Orlando, Florida, United States) p 94124E

[52] Glick S J and Ikejimba L C 2018 Advances in digital and physical anthropomorphic breast phantoms for x-ray imaging *Med. Phys.* **45**

[53] Moloney F, Twomey M, James K, Kavanagh R G, Fama D, O'Neill S, Grey T M, Moore N, Murphy M J, O'Connor O J and Maher M M 2018 A phantom study of the performance of model-based iterative reconstruction in low-dose chest and abdominal CT: When are benefits maximized? *Radiography* **24** 345–51

[54] *Phantom - Triangular array* YouTube uploaded by Anh Le in June 2024 https://www.youtube.com/watch?v=cSf2wJhU9pU

[55] *Import and export file in FreeCAD/GDML* YouTube uploaded by Anh Le in June 2024 https://www.youtube.com/watch?v=LYI3g85vmmQ&t=1s